\documentclass[12pt]{article}
\usepackage[margin=0.5in]{geometry}
\usepackage{color,fancyhdr,framed,lettrine,setspace,listings,array}
\usepackage[T1]{fontenc}   
\usepackage{mathtools,dsfont}
\usepackage{authblk}
\usepackage{braket}
\usepackage{amsthm}
\usepackage{amsfonts}
\usepackage{amssymb}
\usepackage{amsmath,bm}
\usepackage{sansmath}
\usepackage{enumitem}
\usepackage{subeqnarray}
\usepackage{diagbox}
\usepackage{makecell}
\usepackage{times}

\usepackage[utf8]{inputenc}
\newcommand*\diff{\mathop{}\!\mathrm{d}} %Nice looking differential form
\newcommand{\mathbbm}[1]{\text{\usefont{U}{bbm}{m}{n}#1}}
\newcommand{\Id}{\mathbbm{1}} %Identity matrix as seen in school
\DeclarePairedDelimiter\abs{\lvert}{\rvert}%Absolute value of complex number
\DeclareMathOperator{\Tr}{Tr} %Trace operation
\begin{document}
\title{Bi-frequency illumination: a quantum-enhanced protocol}
\author[1,2]{Mateo Casariego}
%\email{mateo.casariego@tecnico.ulisboa.pt}
\author[1,2,3]{Yasser Omar}
\author[4,5,6,7]{Mikel Sanz}
%\email{yasser.omar@lx.it.pt}
%\affil[2]{Physics of Information and Quantum Technologies Group, Instituto de Telecomunicações, Portugal \\Instituto Superior Técnico, Universidade de Lisboa, Portugal}

%\email{mikel.sanz@ehu.es}

\affil[1]{Instituto Superior Técnico, Universidade de Lisboa, Portugal}
\affil[2]{Physics of Information and Quantum Technologies Group, Centro de Física e Engenharia de Materiais Avançados (CeFEMA), Portugal}
\affil[3]{PQI -- Portuguese Quantum Institute, Portugal}
\affil[4]{Department of Physical Chemistry, University of the Basque Country UPV/EHU, Apartado 644, E-48080 Bilbao, Spain}
\affil[5]{EHU Quantum Center, University of the Basque Country UPV/EHU}
\affil[6]{Basque Center for Applied Mathematics (BCAM), Alameda de Mazarredo 14, 48009 Bilbao, Basque Country, Spain}
\affil[7]{IKERBASQUE, Basque Foundation for Science, Plaza Euskadi 5, 48009 Bilbao, Spain}
\date{}
\maketitle

\begin{abstract}
We propose a  quantum-enhanced, idler-free sensing protocol to measure the response of a target object to the frequency of a probe in a noisy and lossy scenario.
{In our protocol, we consider a target with frequency-dependent reflectivity $\eta(\omega)$ embedded in a thermal bath. We aim at estimating the parameter $\lambda = \eta(\omega_2)-\eta(\omega_1)$, since it contains relevant information for different problems. For this, we employ as the resource a bi-frequency quantum state, since it is necessary to capture the relevant information about the parameter.}
Computing the quantum Fisher information $H$ relative to the parameter $\lambda$  in an assumed neighborhood  of $\lambda \sim 0$ for a two-mode squeezed state ($H_Q$), and a pair of coherent states ($H_C$), we show a quantum enhancement in the estimation of $\lambda$. This quantum enhancement grows with the mean reflectivity of the probed object, and is noise-resilient. We  derive explicit formulas for the optimal observables, and propose an experimental scheme based on elementary quantum optical transformations. Furthermore, our work opens the way to applications in both radar and medical imaging, in particular in the microwave domain.
\end{abstract}

\section{Introduction}
Quantum information technologies are opening very promising prospects for faster computation, securer communications, and more precise detection and measuring systems, surpassing the capabilities and limits of classical information technologies \cite{nielsen_chuang_2010, google-quantum-supremacy, PhysRevLett.69.3598, Pirandola_2018, katori}. Namely, in the domain of quantum sensing and metrology \cite{ Giovannetti_2011}, we are currently witnessing a boost of applications to a wide spectrum of physical problems: from gravimetry and geodesy \cite{bongs, menoret, flury, grotti, zeuthen}, gravitational waves \cite{ligo}, clock synchronisation \cite{katori, okeke}, thermometry \cite{thermometry} and bio-sensors \cite{bowen2013, omar, plenio, biosensors, jensen}, to experimental proposals to seek quantum behavior in macroscopic gravity \cite{milburn}, to name just a few. 

While many of the quantum metrology studies focus on unlossy and noiseless (unitary) scenarios, the more realistic, lossy case has also been investigated \cite{lee, changhun, zhang, zhang2, guta2012, davidovich2012, acin2013, sekatski, sekatski2}. Equivalently, one can talk about quantum metrology with open quantum systems. Understanding what are the precision limits of measurements in the presence of loss is a fundamental endeavour in quantum metrology \cite{huelga, friis}. Certain noise properties have been found to be beneficial in some scenarios \cite{huelga2014, sekatski2017}, and quantum error correction schemes have been proposed to overcome decoherence and restore the quantum-enhancement \cite{muschik2017}. 
Quantum illumination (QI) \cite{Lloyd1463, PhysRevLett.101.253601, guha2009, shapiroStory, alsing2019, pirandolaMicrowave, shabir, borre, genovese, gregory, cai2021} is a particularly interesting example of a lossy and noisy protocol where the use of entanglement proves useful even in an entanglement-breaking scenario. QI shows that the detection of a low-reflectivity object in a noisy thermal environment with a low-intensity signal is enhanced when the signal is entangled to an idler that is kept for a future joint measurement with the reflected state. This makes QI a candidate for a quantum radar \cite{lanzagorta}, although a more involved protocol is needed \cite{macconeRadar, Zhuang2021}. The decision problem of whether there is an object or not can be rephrased as a quantum estimation of the object's reflectivity $\eta$,  in order to discriminate an absence ($\eta =0$) from a presence ($\eta \ll 1$) of a low-reflectivity object \cite{Sanz_2017}.

The goal of quantum estimation \cite{holevo, helstrom, caves, doi:10.1142/S0219749909004839, parisBook} is to construct an estimator $\tilde{\lambda}$ for certain parameter $\lambda$ characterizing the system. It is noteworthy that not every parameter in a system corresponds to an observable, and this may imply the need for data post-processing. Either way, the theory provides techniques to obtain an optimal observable --not necessarily unique, \textit{i.e.} whose mean square error is minimal. The estimator $\tilde{\lambda}$ is nothing but a map from the results of measuring the optimal observable to the set of possible values of the parameter $\lambda$. One of the main results of this theory is the quantum Cramér-Rao (qCR) bound, which sets the ultimate precision of any estimator. Whether this bound is achievable or not depends on the data-analysis method used, and on the statistical distribution of the outcomes of different runs of the experiment. In most practical situations, maximum likelihood methods for unbiased estimators, together with a Gaussian distribution of the outcomes of the (independent) runs of the experiment, make the bound achievable. 
In order to find the qCR --and the explicit form of the optimal observable-- one needs to compute the quantum Fisher information \cite{footnote1} (QFI), which roughly speaking quantifies how much information about $\lambda$ can be extracted from the system, provided that an optimal measurement is performed. In general, computing the QFI involves diagonalisation of the density matrix, which makes the obtention of analytical results challenging. However, if one restricts to Gaussian states and Gaussian-preserving operations \cite{parisOlivares, adesso, olivares, serafini}, the so-called symplectic approach simplifies the task considerably \cite{Safranek_2018, _afr_nek_2015, pinel, friis2015, pinel2013, monras, jiang2013, marian2016, gao, nichols, banchi}.
As the QFI is by definition optimized over all POVMs, it only depends on the initial state, often called \textit{probe}. This means that a second optimization of the QFI can be pursued, this time over all possible probes. Moreover, this approach allows us to quantitatively compare different protocols, \textit{e.g.} with and without entanglement in the probe, since an increase in the QFI when the same resources are used --which typically translates into fixing the particle number, or the energy-- directly means an improvement in precision.

{In this article, we propose an idler-free quantum-enhanced, lossy protocol to estimate the reflectivity $\eta(\omega)$ of an object as a function of the frequency when the object is embedded in a noisy environment. In particular, we propose a  method where a bi-frequency state is sent to probe a target --modeled as a beam splitter with a frequency-dependent reflectivity $\eta(\omega)$ and embedded in a thermal environment.  The goal is to obtain an estimator for the parameter $\lambda = \eta(\omega_2)-\eta(\omega_1)$, that captures information about the linear frequency dependence of the object. For simplicity, it is assumed that the frequencies are sufficiently close  so that we can work in a neighborhood of $\lambda\sim 0$.} 

By imposing that the expected photon number is the same in quantum and classical scenarios, we find the QFI ratio between them, and analyze when it is greater than one. We find that the maximum enhancement is obtained for highly reflective targets, and derive explicit limits in the highly noisy case.
We also provide expressions for the optimal observables, proposing a general experimental scheme described in Figure 2 , and motivating applications in microwave technology \cite{Sanz_2018}.  

The article is structured as follows. First, we introduce the model, along with the main concepts and formulas from quantum estimation theory, motivating the use of Gaussian states. Then, we compute the QFI and show the quantum enhancement. Finally, we compute the optimal observables for both the quantum and the classical probes, and briefly discuss applications.

\section{Model, and fundamentals of quantum estimation theory with Gaussian states}

\subsection{Physics of Gaussian states}
When a quantum system has one or more degree of freedom described by operators with a continuous spectrum, we say that the system is a `continuous variable' (CV) system. Within the bosonic CV quantum systems, quantum Gaussian states are defined as the ones arising from Hamiltonians that are at most quadratic in the field operators, which we list in the vector $\hat{\bm{A}}:=(\hat{a}_1, \hat{a}_2, \ldots, \hat{a}_N, \hat{a}^\dagger_1, \hat{a}^\dagger_2, \ldots \hat{a}^\dagger_N$), where $N$ is the number of modes. This ordering of the creation and annihilation operators is commonly referred to as the `complex basis' or `complex form' \cite{adesso}, and allows for a compact way of writing down the commutation relations: $\left[ \hat{\bm{A}}_{{a}}, \hat{\bm{A}}_{{b}}\right]= K_{{a}{b}} \hat{\bm{\text{I}}}$, where ${a, b}=1, \ldots, N$, $\hat{\bm{\text{I}}}$ is the identity operator, and $K=\text{diag}(\Id_N, -\Id_N) $ is a diagonal matrix, $\Id_N$ being the $N\times N$ identity matrix.

 Instead of having to resort to the infinite-dimensional density operator in order to describe a state, Gaussian systems are fully characterized by an $N$-vector called the \textit{displacement vector} and a $N \times N$ matrix, the \textit{covariance matrix}.  We can construct the displacement vector 
\begin{equation}
\bm{d}:=\Tr \left[ \rho \hat{\bm{A}}\right],
 \end{equation} 
 and the covariance matrix 
\begin{equation}
\Sigma := \Tr \left[ \rho \lbrace \Delta \hat{\bm{A}}, \Delta \hat{\bm{A}}^\intercal \rbrace \right],
 \end{equation} 
where $\rho$ is the density operator, $\lbrace \cdot, \cdot \rbrace$ denotes the anticommutator, and $\Delta \hat{\bm{A}}:= \hat{\bm{A}}-\bm{d}$. It is important to bear in mind that other choices of basis lead to different, but equivalent definitions. In fact, in the following sections we will start by writing down covariance matrices in the so-called `quadrature basis' $\left( \hat{x}_1, \ldots, \hat{x}_N, \hat{p}_1, \ldots, \hat{p}_N\right)$ with the \textit{canonical} position and momentum operators  defined by the choice $\kappa_1=2^{-1/2}$ in $\hat{a}_k = \kappa_1(\hat{x}_k + i \hat{p}_k)$\footnote{The `quantum optics' convention takes $\kappa_1=1$. }. A key result with important consequences in the context of Gaussian states is the normal mode decomposition \cite{ arvind, simon}, which follows the more general theorem due to Williamson \cite{williamson} and that, from a physical point of view, establishes that any Gaussian Hamiltonian (i.e., quadratic) is equivalent --up to a unitary-- to a set of free, non-coupled harmonic oscillators.  This apparent simplicity of Gaussian states, however, has a rich structure when it comes to analyzing their Hilbert space properties, as well as information-theoretic quantities such as the quantum Fisher information, entropies, and so on.  We can state the result in the following way: any positive-definite Hermitian matrix $\Sigma$ of size $2N\times 2N$ can be diagonalized with a symplectic matrix $S$: $\Sigma = S D S^\dagger$, where $D = \text{diag}\left( \nu_1,  \ldots, \nu_N, \nu_1, \ldots, \nu_N\right)$ with $\nu_{a}$ the symplectic eigenvalues of $\Sigma$, that are the positive eigenvalues of matrix $K\Sigma$. An important result for what follows is that a state is pure if and only if all the symplectic eigenvalues are one: $\nu_{a}=1$ $\forall {a}$, and $\nu_{a} \geq 1$ for any Gaussian state.

\subsection{Quantum estimation}
Quantum metrology is so related to quantum estimation that sometimes the two terms are used as synonyms. Incidentally, quantum sensing could be seen as a quantum estimation or metrology problem that deals with a binary question: is the value of the parameter of interest zero or not? Terminology aside, quantum estimation deals with the problem of measuring things that may not be encoded in observables \textit{per se}, i.e., it allows for the obtention of measurable quantities that do not necessarily correspond to linear functions of the density matrix, and it teaches us what the ultimate precision limits are, whether one can attain them or not, and how to attain them. The most common approach to attack the problem of metrology is from the notion of classical \textit{frequentist} estimation. Although the --perhaps more realistic-- approach of Bayesian quantum estimation theory \cite{personick, marcin, pi, morelli, rubio1, rubio2, rubio3} exists and is a rich field of research, here we will take the former,  tacitly assuming a \textit{local} estimation strategy {\cite{kay1993, demko2015}, that happens when there is some prior knowledge of the interval where the true value of the parameter (or parameters) of interest may lie, hence its name: the true parameter value is localized into some interval rather than completely unknown (in this case, the estimation is called  \textit{global}).} In the local approach, the quantum Fisher information (QFI) matrix emerges as the figure of merit for the quantification of the maximum amount of information one can extract from the system. 

While classical parameter estimation deals only with the statistics of measurement outcomes, and answers questions of attainability in the presence of statistical noise (with various properties that can affect the scaling of the precision with which one estimates the parameter), quantum estimation addresses the problem of \textit{what} to measure, and imposes additional limits to the precision due to to the fundamental probabilistic nature of quantum mechanics. 
Indeed, the quantum Fisher information (QFI) matrix,  can be seen as an optimization of the classical Fisher information --a measure for the amount of information relative to a set of parameters $\bm{\lambda}$ a system contains-- over all possible measurements, or POVMs.

The QFI can be interpreted geometrically by means of a notion of distance in the Hilbert space spanned by density operators. Among the many candidates, the Bures distance
\begin{equation}
D^2_B(\rho_1, \rho_2) := 2\left( 1-\sqrt{F(\rho_1, \rho_2)}\right),
\end{equation}
{where $F(\rho_1, \rho_2) := \left(\Tr \left[ \sqrt{\sqrt{\rho_1}\rho_2\sqrt{\rho_1}}\right]\right)^2$ is the Uhlmann fidelity} between states $\rho_1$ and $\rho_2$, the one correctly linking estimation to geometry. This makes the interpretation of quantum estimation straightforward: it depends upon the distinguishability between states. If $\bm{\lambda}$ is a vector of parameters that defines a (possibly continuous) family of states $\lbrace \rho_{\bm{\lambda}}\rbrace$ , then the Bures distance between two infinitesimally close states can be related to a metric tensor, which is no other than the QFI matrix:
\begin{equation}
D^2_B(\rho_{\bm{\lambda}}, \rho_{\bm{\lambda} + \diff \bm{\lambda} }) = \frac{H_{ab}(\bm{\lambda})}{4}.
\end{equation}
A large QFI translates in a large distinguishability between states. In this paper we will focus on the single parameter case, for which the QFI is a scalar that can be computed using the following basis-dependent formula 
\begin{equation}\label{eq:qfi}
H(\lambda)= 2\sum_{m,n} \frac{\abs{\bra{\Phi_{m}}\partial_\lambda {\rho}_\lambda \ket{\Phi_{n}}}^2}{\rho_{m}+\rho_{n}},
\end{equation}
where $\lbrace \rho_{m}, \ket{\Phi_{m}} \rbrace$ are the eigensolutions to ${\rho}_\lambda \ket{\Phi_{m}} = \rho_{m} \ket{\Phi_{m}}$, and ${\rho}_\lambda$ is the measured, or received state.
Moreover, the theory also provides a way of finding an optimal observable, whose outcomes allow us to construct an estimator \cite{doi:10.1142/S0219749909004839}:
\begin{equation}\label{eq:opt-obs}
\hat{O}_\lambda = \lambda \Id + \frac{\hat{L}_\lambda}{H(\lambda)},
\end{equation}
 where $\hat{L}_\lambda$ is a symmetric logarithmic derivative (SLD) that solves the equation $\lbrace\hat{L}_\lambda, {\rho}_\lambda\rbrace = 2\partial_\lambda {\rho}_\lambda$, where $\lbrace \cdot, \cdot \rbrace$ is the anticommutator. When the estimator $\tilde{\lambda}$ is constructed using a maximum likelihood methid, the so-called quantum Cramér-Rao bound (qCRB) \cite{cramer, rao} is  asymptotically achieved, meaning that the observable in \eqref{eq:opt-obs} has the smallest possible variance:
\begin{equation}
\text{var}(\hat{O}_{\bm{\lambda}}) \geq  \frac{1}{M H(\bm{\lambda})},
\end{equation}
where {$\text{var}(\hat{O}) := \langle \hat{O}^2 \rangle - \langle \hat{O} \rangle^2$} denotes the variance of operator $\hat{O}$, $M$ is the number of repetitions, and $H(\bm{\lambda})$ is the QFI.

The problem, however, can become mathematically challenging due to the diagonalization implicit in \eqref{eq:qfi}. In the next section we review some results that help us circumvent these issues, as long as we stick to Gaussian states.
\subsubsection{Gaussian quantum estimation}
As shown in \cite{Safranek_2018}, when we are in the presence of Gaussian states and Gaussian-preserving channels, there is no need to diagonalize the density matrix in Eq. \eqref{eq:qfi} in order to find the QFI. For a single parameter, the QFI can be computed using
\begin{align}\label{eq:QFI}
\begin{split}
H(\lambda) &= \frac{1}{2(\det[A]-1)}\left[
\det[A] \Tr\left[ (A^{-1}\partial_{\lambda}A)^2\right] 
+ \sqrt{\det[\Id_2 + A^2]}\Tr\left[\left((\Id_2 + A^2)^{-1}\partial_{\lambda}A\right)^2 \right]  
 \right. \\
&-\left. 4\left( \nu_+^2 - \nu_-^2 \right) \left(\frac{(\partial_\lambda \nu_+)^2}{\nu_+^4-1} - \frac{(\partial_\lambda \nu_-)^2}{\nu_-^4-1}\right)\,\right]+ 2\partial_\lambda \bm{d\,}^\dagger\Sigma_\lambda^{-1}\partial_\lambda \bm{d},
\end{split}
\end{align}
where the \textit{dot} over $A$ and $\vec{d}$ denotes  derivative with respect to $\lambda$, and $\nu_{\pm}$ are the \textit{symplectic eigenvalues} of ${\Sigma}_\lambda$, defined following Ref.  \cite{_afr_nek_2015}
\begin{equation}\label{eq:symp-eigenv}
2\nu_{\pm}^2 := \Tr[A^2] \pm \sqrt{\left(\Tr[A^2]\right)^2-16 \det[A]},
\end{equation}
with the matrix $A$  given by 
$A := i {\Omega}{T}{\Sigma}_\lambda{T}^\intercal$,  $\Omega:=\text{antidiag}(\Id_2, -\Id_2)$, and ${T}_{ij}:=\delta _{j+4,2 i}+\delta _{j,2 i-1}$ is the matrix that changes the basis to the \textit{quadrature basis}
$(\hat{x}^{\text{th}}_1,  \hat{x}^{\text{S}}_1, \hat{x}^{\text{th}}_2, \hat{x}^{\text{S}}_2, \hat{p}^{\text{th}}_1, \hat{p}^{\text{S}}_1, \hat{p}^{\text{th}}_2, \hat{p}^{\text{S}}_2)^{\intercal}$. 
For a Gaussian state $\left( {\Sigma}_\lambda, \vec{d}_\lambda\right)$ written in the complex basis,  the symmetric logarithmic derivative in Eq. \eqref{eq:opt-obs} can be obtained as in  Ref. \cite{Safranek_2018}: 
\begin{equation}\label{eq:sld}
\hat{L}_\lambda = \Delta \vec{\hat{A}}^\dagger {\mathcal{A}}_\lambda  \Delta \vec{\hat{A}} -\Tr [{\Sigma_\lambda}{\mathcal{A}}_\lambda]/2 + 2\Delta \vec{\hat{A}}^\dagger {\Sigma}_\lambda^{-1}\partial_\lambda\vec{d}_\lambda,
\end{equation}
where $\Delta \vec{\hat{A}} := \vec{\hat{A}} - \vec{d}_\lambda$,
 $\vec{\hat{A}}$  the \textit{complex basis} vector of bosonic operators, ${\mathcal{A}}_\lambda:= {\mathcal{M}}^{-1}\partial_\lambda \vec{d}_\lambda$, where   ${\mathcal{M}} =\bar{ {\Sigma}}_\lambda \otimes {\Sigma}_\lambda - {K}\otimes {K}$, where the bar denotes complex conjugate, and ${K} := \text{diag}\left(\Id_2, -\Id_2\right)$.
Note that when $\lambda\rightarrow 0$ we have $\hat{O}_{\lambda=0}\equiv \hat{O} = \hat{L}_{\lambda=0}/H(\lambda=0)$, since both limits exist independently. This limit is of our interest because we will work in a neighbourhood of $\lambda \sim 0$, \textit{i.e.} the measured value of the parameter is expected to be small (\textit{i.e.} we shall adopt a local estimation strategy).

\subsection{Model}
\begin{figure}[ht]
\centering
\includegraphics[width=0.4\linewidth]{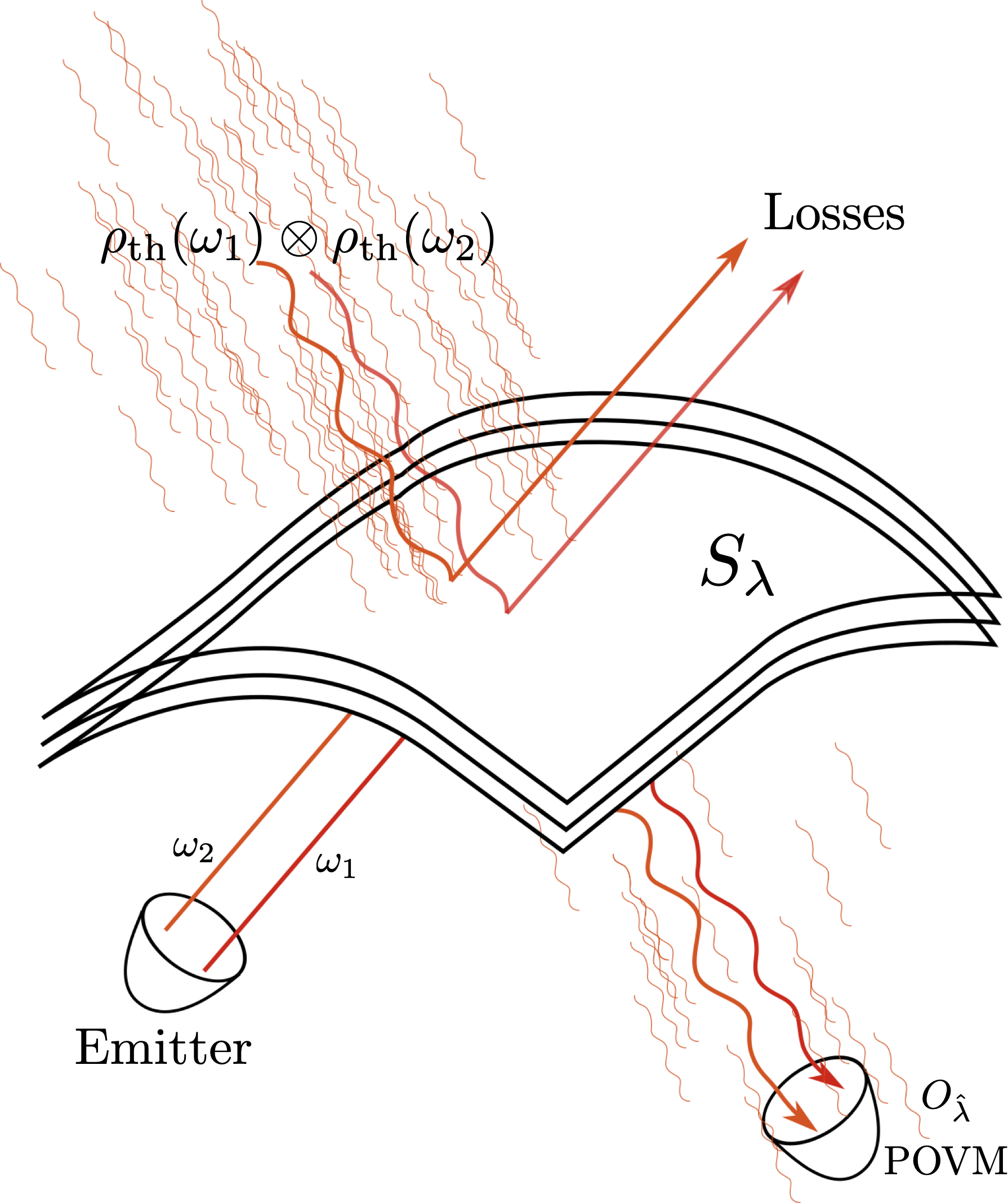}
\caption{An object reflects a bi-frequency beam (notice the similar but different colors of the two beams coming out of the emitter), and mixes it with a thermal bath for each frequency with the same expected photon number, coming from the upper left. The transmitted signal and reflected thermal state are lost, and a measurement $O_{\hat{\lambda}}$ is performed onto the available part (lower right corner), whose expectation values converge, after classical data processing, to an estimator of the parameter $\lambda$ encoded in the object. In our case, $S_\lambda$ represents the transformation associated to a multi-layered object, modeled as a beam splitter, where $\lambda := \eta_2 - \eta_1$ is the parameter to be estimated, where $\eta_i =\eta(\omega_i)$ are the reflectivities for the different frequencies. The emitter can produce either a pair of coherent states (classical strategy) or an entangled, two-mode squeezed state. The latter proves advantageous in the parameter estimation, giving a strictly larger quantum Fisher information.}
\label{fig:diag}
\end{figure}

The model is synthesized in Figure \ref{fig:diag}: the target object, modeled as a beam splitter with a frequency-dependent reflectivity is subject to an illumination with a bi-frequency probe. The transmitted signal is lost, and a only the reflected part is collected for measurement. For a single frequency, a beam splitter is characterised by a unitary operator
\begin{equation}\label{eq:BS}
{U}({\omega})
\equiv   
\exp
\left[
\arcsin\left(\sqrt{\eta(\omega)}\right)(
\hat{s}_\omega^\dagger \hat{b}_\omega e^{i\varphi} - \hat{s}_\omega \hat{b}_\omega^\dagger e^{-i\varphi}
)
\right],
\end{equation}
where $\eta(\omega)$ is a frequency-dependent reflectivity, related to transmittivity $\tau$ via $\eta(\omega) + \tau(\omega) = 1$.

We assume for simplicity that $\varphi = 0$, \textit{i.e.} there is no phase difference between transmitted and reflected signals.
This unitary maps states (density matrices) that live in the density matrix space associated with Hilbert space $\mathcal{H}$, $\mathcal{D}(\mathcal{H})$ to itself.
Formulating the problem from a density operator perspective, we have that the received state is
\begin{equation}\label{eq:rholambda}
{\rho}_\lambda = 
\Tr_{S_1} \Tr_{S_2}
\left[{U}_\lambda{\rho}{U}^\dagger_\lambda
\right],
\end{equation}
where the parameter is defined as
$\lambda = \eta(\omega_2)-\eta(\omega_1)$, and ${\rho}\in \mathcal{H}_{S_1, S_2, B_1, B_2}$ is a four-mode state that includes the two signals (the two-mode state that we can control) and two thermal environments of the form ${\rho}^{th}_1\otimes {\rho}^{th}_2$, where the subscript indicates the frequency, \textit{i.e.}
${\rho}^{th}_a = (1+N_{\text{th}})^{-1}\sum_{n=0}^\infty (N_{\text{th}}/(1+N_{\text{th}}))^n
\ket{n}_a\bra{n}$
	where $N_{\text{th}} = \Tr ({\rho}^{th}_a \hat{b}^\dagger_{\omega_a}\hat{b}_{\omega_a})$ is the average number of thermal photons, which we assume to be the same for the two modes. 
Note that in order to obtain the explicit form of the interaction ${U}_\lambda$ in Equation \eqref{eq:rholambda} one just needs to reparametrize the four-mode unitary ${U}(\omega_1)\otimes{U}(\omega_2)$ using the difference of reflectivities $\lambda \equiv \eta_2-\eta_1$ and Equation \eqref{eq:BS}.	
The equal thermal photon number is an accurate approximation as long as the frequency difference $\Delta\omega \equiv\omega_2 - \omega_1$ is sufficiently small. To make this statement more quantitative, let us assume two different thermal photon densities, $N_1$ and $N_2$. The Bose-Einstein distribution for photons is
$N_i  \propto 1/(e^{\beta \omega_i }-1)$ where $\beta \equiv \hbar /k_B T$ is a function of the temperature $T$. Then,
\begin{equation}
\frac{N_1}{N_2} = \frac{e^{\beta \omega_1 }-1}{e^{\beta \omega_2 }-1} = \frac{1}{1 + \frac{\beta \Delta\omega e^{\beta \omega_1}}{e^{\beta \omega_1}-1}},
\end{equation}
we see that up to first order in $\beta \Delta\omega$, the last expression reduces to $1 -\Delta\omega / \omega_1$. This means that $N_1 \approx N_2$ if $\Delta\omega / \omega_1 \ll 1$. In particular, for $T= 300 \text{ K}$ and $\omega_1/2\pi = 5 \text{ GHz}$ the expected thermal photon number is roughly 1250. It is straightforward to check that for these frequencies and temperatures, the above approximations are good (\textit{i.e.} $\sim 4$\% of relative error) for frequency differences up to 20\%.

{
Because we are working within the local estimation approach and our goal is to find observables that saturate the qCRB, we shall take the true value of $\lambda$ to be  exactly zero. This means that the goal of the protocol is to increase one's confidence about this initial ansatz of the parameter being zero, and be able to tell when it is close but not exactly zero. Hence, we work in a neighborhood of $\lambda \sim 0$ --~which can be implemented by taking the limit $\lambda \rightarrow 0$ in the derived expressions. Moreover, this relies on a physical assumption, since we are interested in probing regions of  $\eta(\omega)$ that do not change drastically, \textit{i.e.} that are well approximated by a linear function with either no slope or a small one.} In this sense, the protocol is a quantum sensing one, since we are interested in answering the question of whether the parameter either vanishes or is small.

%Completely positive, trace preserving (CPTP) Gaussian quantum channels are easily characterized by two matrices $X$ and $Y$. If an initial multipartite state has first and second moments given by $\bm{d}_\text{in} =(\bm{d}_A, \bm{d}_B)^\intercal$ and $$\Sigma_\text{in} = \left[\begin{array}{cc}
%  \Sigma_A & \varepsilon_{AB}\\
% 	\varepsilon_{AB}^\intercal &  \Sigma_B,
%\end{array}\right] $$ then the action of a Gaussian CPTP map that affects only subsystem $A$ is
% $\bm{d}_\text{in} \mapsto \bm{d}_\text{in} = (X \bm{d}_A, \bm{d}_B)^\intercal$ and $$\Sigma_\text{in} \mapsto \Sigma_\text{out} = \left[\begin{array}{cc}
%  X\Sigma_AX^\intercal + Y & X\varepsilon_{AB}\\
% 	\varepsilon_{AB}^\intercal  X^\intercal&  \Sigma_B\end{array}\right] $$.
 {It is also worth discussing briefly the effect of absorption loss due to the medium through which the signal travels. These  can be accommodated in the model by means of an additional beam splitter. The medium through which the signal travels can be seen as an array of infinitesimal beam splitters, each of which having the same reflectivity, and mixing some incoming signal with the same thermal state. For a travel distance $L$, the flying mode will see a reflectivity
\begin{equation}
\eta_{\text{abs}} = 1-e^{-\mu L},
\end{equation}
where $\mu$ is a parameter characterizing the photon-loss of the medium. A concatenation of beam splitters can be easily put into a single one, as long as they are embedded in the same environment, which is our case. For beam splitters of transmittivities $\tau_1$ and $\tau_2$ the resulting transmittivity of them combined is simply the product: $\tau = \tau_1 \tau_2$. Thus, acommodating absorption losses into our model is trivially obtained by the transformation $\tau \mapsto e^{-\mu L} \tau $. Since the QFI deals with derivatives with respect to the parameter to be estimated, and ultimately we are interested in QFI  \textit{ratios} between a quantum protocol and its classical counterpart, the above transformation will not affect the overall results, since multiplicative factors will cancel out.}
\section{Results: Quantum Fisher information}
In this section we compute the QFI for two different probes: an entangled two-mode squeezed (TMS) state, and a pair of coherent beams. {The choice of the TMS state over other possible entangled states is motivated by the fact that these are customarily produced in labs, both in optical --e.g. with non-linear crystals, and in microwave frequencies --using Josephson parametric amplifiers (JPAs).}
\subsection{Two-mode squeezed vacuum state}
The TMS vacuum (TMSV) state is the continuous-variable equivalent of the Bell state, being the Gaussian state that optimally transforms classical resources (light, or photons) into quantum correlations. The TMSV state is a cornerstone in  experiments with quantum microwaves \cite{PhysRevLett.107.113601, PhysRevLett.109.183901, RevModPhys.77.513, casariego2022, gonzalezraya2022}.  In our case, we are interested in states produced via nondegenerate parametric amplification, in order to have two distinguishable frequencies.
The state can be formally written as:
$\ket{\psi}_{\text{TMSV}}:=(\cosh r)^{-1}\sum_{n=0}^\infty \left(
-e^{i\phi}\tanh r \right)^n \ket{n,n}$,
where $r \in \mathbb{R}_{\geq 0}$ is the \textit{squeezing parameter}.  For simplicity we take $\phi=0$. {In any realistic application, the TMSV state should be replaced by a TMS thermal state, which can be defined as the one obtained by applying the two-mode squeezing operation to a pair of uncorrelated thermal states $\rho_\text{th, 1}$, and $\rho_\text{th, 2}$ with mean thermal photon numbers $n_1$ and $n_2$, respectively, and hence resulting in a mixed state \cite{TMST}. The expected total photon number in these states is given by $N_\text{TMST}=\braket{\hat{N}_1+\hat{N}_2} = (n_1+n_2)\cosh 2r + 2 \sinh ^2 r$, where $ \hat{N}_i \equiv \hat{a}^\dagger_{S_i} \hat{a}_{S_i}$ for $i=1,2$.
Typically, one has $n_1 = n_2 \equiv n$, which gives us a symmetric TMST state. In this case we define the signal photon number $N_\text{S}$ as the photon number in each of the modes,  $ N_{\text{S}} \equiv N_\text{TMST} /2 = n(1+2{N_r})+{N_r}$, where $N_r \equiv \sinh^2 r$ \cite{footnote2}. In microwaves, a squeezing level $S = -10 \log_{10}\left[(1+2 n)\exp\left(-2r\right)\right]$ of 9.1 dB has been reported \cite{finite-time} for $n=0.34$ and $r\sim 1.3$, using JPAs operating at roughly 5 GHz with a filter bandwidth of 430 kHz. This corresponds to $N_\text{S} \sim 8$.}

The total initial (real) covariance matrix --written in the \textit{real basis} $(\hat{x}^{\text{th}}_1, \hat{p}^{\text{th}}_1, \hat{x}^{\text{S}}_1, \hat{p}^{\text{S}}_1,\hat{x}^{\text{th}}_2, \hat{p}^{\text{th}}_2,\hat{x}^{\text{S}}_2, \hat{p}^{\text{S}}_2)^{\intercal}$-- is given by

{
\begin{equation}
{\Sigma} = 
N\left(
\begin{array}{cccc}
 N^{-1}{\Sigma}_{\text{th}} & 0 & 0 & 0\\
 0 & {\Sigma}_{r} & 0 & {\varepsilon}_r \\
 0 & 0 & N^{-1}{\Sigma}_{\text{th}} & 0\\
 0 & {\varepsilon}^{\intercal}_r & 0 &{\Sigma}_{r}
\end{array}
\right),
\end{equation}
where $N\equiv 1+2n$, ${\Sigma}_{\text{th}} = (1+2N_{\text{th}})\Id_2$ is the real covariance matrix of a thermal state, ${\Sigma}_{r} = \cosh(2r)\Id_2$ corresponds to the diagonal part of one of the modes in a TMSV state, and ${\varepsilon}_r = \sinh(2r)\sigma_Z$ is the correlation between the two modes, where $\sigma_Z$ is the Z Pauli matrix. Note that the covariance matrix of the thermal TMS state is simply $N$ times the one of the TMSV state.

The displacement vector of a TMST state is identically zero $\bm{d}_{\text{TMST}}=\bm{0}$, so the last term of Eq. \eqref{eq:QFI} vanishes.
Under the assumption that the object does not entangle the two modes,  we have that the symplectic transformation is  ${S}(\eta_1, \eta_2)  = {S}_{\text{BS}}(\eta_1) \oplus {S}_{\text{BS}}(\eta_2)$ \cite{footnote3}, where 
\begin{equation}
{S}_{\text{BS}}(x) = 
\left(
\begin{array}{cc}
 \sqrt{x}\Id_2 & \sqrt{1-x}\Id_2 \\
 -\sqrt{1-x}\Id_2 & \sqrt{x}\Id_2
\end{array}
\right)
\end{equation}
is the real symplectic transformation associated with a beam splitter of reflectivity $x$. We define the parameter of interest as $\lambda \equiv \eta_2 - \eta_1$.
With this, ${S}(\eta_1, \eta_2)$ becomes a function of $\lambda$. For simplicity, we define ${S}_\lambda := {S}(\eta_1,  \eta_1 + \lambda)$.
The full state after the signals get mixed with the thermal noise is given by 
$\tilde{{\Sigma}}_\lambda \equiv {S}_\lambda {\Sigma} {S}_\lambda^\intercal$.}
In covariance matrix formalism, partial traces are implemented by removing the corresponding rows and columns \cite{adesso}; in our case the rows and columns 1, 2, 5, and 6. The resulting \textit{received} covariance matrix reads as follows
\begin{equation}
{\Sigma}_\lambda
=
\left(
\begin{array}{cc}
a  \Id  & b \sigma_Z \\
b \sigma_Z & c \Id
\end{array}
\right),
\end{equation}
{with $a \equiv 1+ 2 N_\text{th}+2\eta _1 (2 N_{r} +4 nN_{r}  - N_\text{th})$, $b \equiv 2(1+2n) \sqrt{2\eta _1N_\text{S}(\eta _1+\lambda ) (2 N_\text{r}+1)} $, and $c\equiv  (1+2 n)\left(1+4 \lambda  N_{r}+\eta _1 (4 N_{r}-2 N_\text{th})+2 (1-\lambda) N_\text{th}\right)$.

For this state, the symplectic eigenvalues $\nu_{\pm}$ defined in Eq. \eqref{eq:symp-eigenv} are strictly larger than one for any value of the parameters $n$, $N_r$, $N_\text{th}$, and $\eta_1$, other than $\eta_1=1 \land N_\text{th}=0$ so there is no need of any regularization scheme  \cite{_afr_nek_2015}. Indeed, this is due to the mixedness of the received state: regularization is only needed for pure states.

We obtain the function $H_Q (\lambda)$ from Eq. \eqref{eq:QFI}, and compute the two-sided limit $H_Q\equiv \lim_{\lambda \rightarrow 0}H_Q (\lambda)$ when the parameter $\lambda$ goes to zero, finding
\begin{equation}
\begin{split}
H_Q =\kappa &\left[\eta _1 \bar{n} \left(\bar{N}_\text{th} (4 n N_r+n+4 N_r-2 N_\text{th})-2 \eta _1 (2 (n+1) N_r \bar{N}_\text{th}+N_\text{th} (n-N_\text{th}))\right) \right.\\
&\left. +\beta\right) \left(\eta _1 \bar{n} \left(\bar{N}_\text{th} (4 n N_r+n+4 N_r-2 N_\text{th})-2 \eta _1 (2 (n+1) N_r \bar{N}_\text{th}+N_\text{th} (n-N_\text{th}))\right) \right.\\
&\left.+\beta+1\right]
\end{split}
\end{equation}
where
\begin{equation}
\begin{split}
\kappa^{-1}\equiv \bar{n}^2 &\left[2 \eta _1 (\bar{N}_\text{th} (N_r (\bar{n} (8 (n+1) N_r^2+6 n N_r+n)-4 N_r)+2 N_\text{th}^2 (4 n N_r+n+6 N_r)\right.\\
&-\left. 2 N_r N_\text{th} (2 (6 n+5) N_r+2 n-1)-2 N_\text{th}^3)-2 \eta _1 (-N_\text{th} (4 (n+1) N_r+n)\right.\\
&+\left.N_r ((4 n+2) N_r-1)+N_\text{th}^2) (2 (n+1) N_r \bar{N}_\text{th}+N_\text{th} (n-N_\text{th})))\right.\\
&\left.+2 n (2 N_r+1) N_r \bar{N}_\text{th}^2+4 N_r^2 (6 N_\text{th} (N_\text{th}+1)+1)-4 N_r N_\text{th}^2 (4 N_\text{th}+3)+N_\text{th}^2 \bar{N}_\text{th}^2\right]
\end{split},
\end{equation}

and  $\bar{N}_\text{th} \equiv 1+2 N_\text{th}$,   $\bar{n}\equiv 1+2n$, and $\beta \equiv n \bar{N}_\text{th}^2+2 N_\text{th} (N_\text{th}+1)$.

}

\subsection{Coherent states}
{
Here we use a pair of coherent states as probe: $\ket{\psi} =\ket{\alpha}\otimes\ket{\alpha}$. The total expected photon number in this state is $2N_\text{C}:= 2\left| \alpha^2\right|$. For simplicity we take $\alpha \in \mathbb{R}$. Moreover, since we will compare with the TMST state, we set $\alpha^2 = n(1+2N_r)+N_r$.
The initial covariance matrix is simply given by the direct sum of two identity matrices (corresponding to each of the coherent states), and two thermal states. After the interaction and the losses, the measured covariance matrix is 

\begin{equation}
{\Sigma}_\lambda
=
\left(
\begin{array}{cc}
d \Id  &0 \\
0 & f \Id
\end{array}
\right),
\end{equation}
where $d =1+ 2 N_\text{th}\tau _1 $, $f = 1+ 2 N_\text{th}(\tau _1-\lambda)$.

The initial displacement vector in the real basis is $\bm{d}_{0}^\intercal = (0,0,\sqrt{2}\alpha,0,0,0,\sqrt{2}\alpha,0)$ which leads --after the interaction and the trace of the losses-- to  $\bm{d}^\intercal = \alpha(\sqrt{2\eta_1},0, \sqrt{2(\eta_1 + \lambda)})$.
 The symplectic eigenvalues are also larger than one here. 
Inserting these in Eq. \eqref{eq:QFI}, and taking the limit $\lambda\rightarrow 0 $, we find that the QFI for the coherent state is
\begin{equation}
H_C = \frac{4 N_\text{th}^2 \left((1+2 N_\text{th} \tau_1)^2+1\right)}{(1+2 N_\text{th} \tau_1)^4-1}+\frac{n(1+2N_r)+N_r	}{\eta_1(1+2 N_\text{th} \tau_1)},
\end{equation}
where $\tau_1 = 1-\eta_1$ is the transmittivity. Having computed both the quantum and the classical QFIs, in the next section we analyze their ratio $H_Q / H_C$, a quantifier for the quantum enhancement.}

\subsection{Comparison: Quantum enhancement}
\begin{figure*}
\centering
\includegraphics[width=\linewidth]{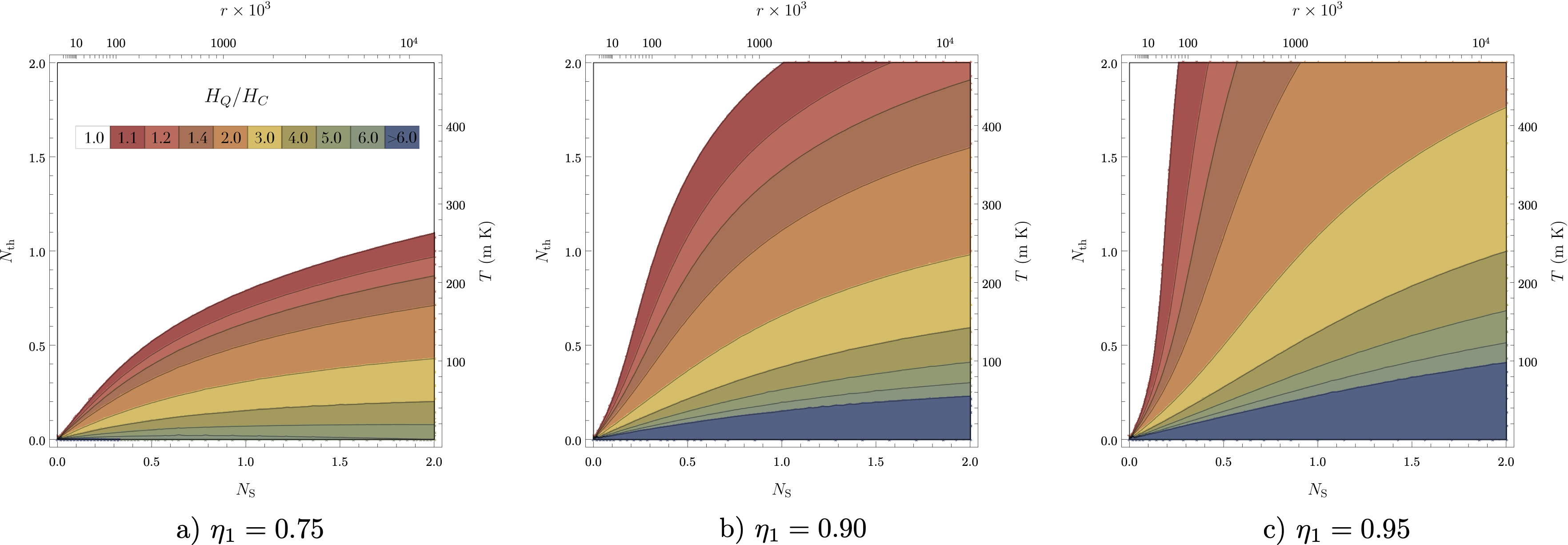}
\caption{Values, represented by a non-linear color grading, of the quantum enhancement given by the ratio $H_Q/H_C$ of the quantum Fisher information of the two-mode squeezed vacuum state probe $H_Q$ by the quantum Fisher information of the coherent states probe $H_Q$ as a function of the photon numbers of the signal ($N_\text{S}$) and of the thermal bath ($N_\text{th}$), for a reflectivity $\eta_1$ of a) 0.75, b) 0.90, and c) 0.95.
 Equivalently, scales of squeezing, $r$, given by $\sqrt{N_{\text{S}}}=\sinh r$, and temperature $T$ in Kelvin, are provided. The relation between temperature and mean thermal photon number is obtained via the usual Bose-Einstein distribution $N_\text{th} = 1/(\exp(E/k_B T) - 1)$ when the energy is set to $E=\hbar \omega = h \nu$, which requires a choice of the frequency $\nu$. We have taken $\nu = 5$ GHz, a typical frequency of microwaves.
White represents no quantum enhancement, \textit{i.e.} $H_Q / H_C =1$. We clearly see that as $\eta_1$ grows, the quantum enhancement becomes not only more significant, but also easier to achieve with less signal photons. Importantly, as the reference reflectivity $\eta_1$ grows,  the protocol becomes more resilient to thermal noise.}
\label{fig:QFIratios}
\end{figure*}

We analyze the ratio between the TMST state's QFI ($H_Q$) and the coherent pair's QFI ($H_C$) for different situations. As a first approximation and to simplify the discussion, we take the limit where $n\rightarrow 0$, which corresponds to a TMSV state input.  Finding values of $(\eta_1, N_\text{th}, N_\text{S})$  such that the ratio $H_Q / H_C$ is larger than one means that one can extract more information about parameter $\lambda$ using a TMST state than using a coherent pair, provided an optimal measurement is performed in both cases. In Figure \ref{fig:QFIratios} we plot the results for various values of $\eta_1$.
We can immediately see that the ratio gets larger  for large values of $\eta_1$, \textit{i.e.} for highly reflective materials. In particular, we find the high-reflectivity limit the ratio converges even when the individual QFIs do not (since they correspond to a pure state being transmitted):
\begin{equation}
\lim_{\eta_1 \rightarrow 1}\frac{H_\text{Q}}{H_\text{C}} = \frac{N_\text{S}^2 \left(8 N_\text{th} (N_\text{th}+1)+4\right)+4 N_\text{S} N_\text{th}^2+N_\text{th}^2}{N_\text{th} \left(N_\text{S} (4 N_\text{th}+2)+N_\text{th}\right)}
\end{equation}
which converges to 
$1 + 8 N_\text{S}^2/(4N_\text{S}+1)$
in the highly noisy scenario $N_\text{th}\gg 1$. Using a squeezing of $r\sim 1.3$ which is experimentally realistic for microwave quantum states, and that corresponds to an expected photon number of $N_\text{S}\sim2.9$, we expect  to find a quantum-enhancement of roughly a factor of 6, \textit{i.e.}, $ H_\text{Q}/H_\text{C} \sim 6.4$ in the highly reflective limit.

In the next section we explicitly compute the observables that lead to an optimal extraction of $\lambda$'s value for both the classical and the quantum probes.

\section{Optimal observables}
Here we address the question of how to extract the maximum information about parameter $\lambda$ for each of the probes. The theory provides us with explicit ways to compute an optimal POVM, which despite being not unique, provides us with an optimal measurement strategy: upon measuring the outcomes and possibly after some classical data-processing, the results asymptotically tend towards the true value of the parameter to be estimated.
\subsection{Optimal observable for the TMSV state probe}
 Computing the SLD in Eq. \eqref{eq:QFI} and inserting it in Eq. \eqref{eq:opt-obs} we find 
\begin{equation}\label{eq:QObserv}
\hat{O}_Q = L_{11}\hat{a}_1^\dagger \hat{a}_1 + L_{22}\hat{a}_2^\dagger \hat{a}_2 + L_{12}\left( \hat{a}_1^\dagger \hat{a}_2^\dagger + \hat{a}_1 \hat{a}_2\right) + L_0 \Id_{12},
\end{equation}
where the general expressions  for the coefficients can be found in Appendix \ref{app:A}.
{ The variance of this operator is found to be $\text{var}\left(\hat{O}_Q\right) = 2N_\text{S}^2L_{12}\left(1+N_\text{S}\right)$. We can numerically test the validity of the qCR bound for this observable by examining the bound itself for the extreme choice of $M=1$. The saturation of the bound produces the following relation:
\begin{equation}
\text{var}\left(\hat{O}_Q\right) H_Q = 1.
\end{equation}
Now, as the left hand side a function of $(N_\text{S}, N_\text{th}, \eta_1)$, we can give different values to the reflectivity and find the limiting condition between $N_\text{S}$ and $N_\text{th}$, which is depicted in FIG. \ref{fig:qCR}. Naturally, the larger $M$, the better results we can achieve, but $M=1$ proves the existence of a choice of parameters for which the bound is saturated. 
\begin{figure}[ht]
\centering
\includegraphics[width=0.6\textwidth]{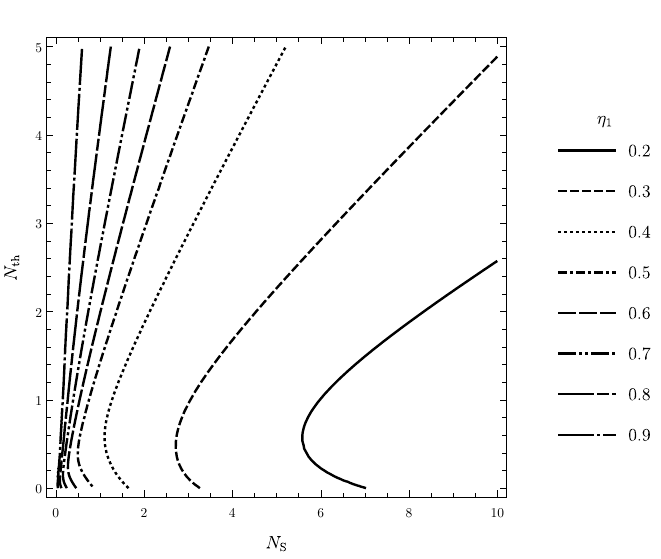}
\caption{ Proof of the saturation of the quantum Cramér-Rao bound for the optimal observable $\hat{O}_Q$ given in Eq.~\eqref{eq:QObserv} for different values of the reflectivity $\eta_1$, expressed as the existence of a real function $N_\text{th} =  N_\text{th}(N_\text{S})$, for the extreme case of just one experimental run ($M=1$). As the reflectivity grows, we observe an interesting behavior: the best choice of $N_\text{th}$ --defined as the one that saturates the bound while keeping $N_\text{S}$ as low as possible-- is actually non-vanishing.}
\label{fig:qCR}
\end{figure}
}

\begin{figure}[ht]
\centering
\includegraphics[scale=.15]{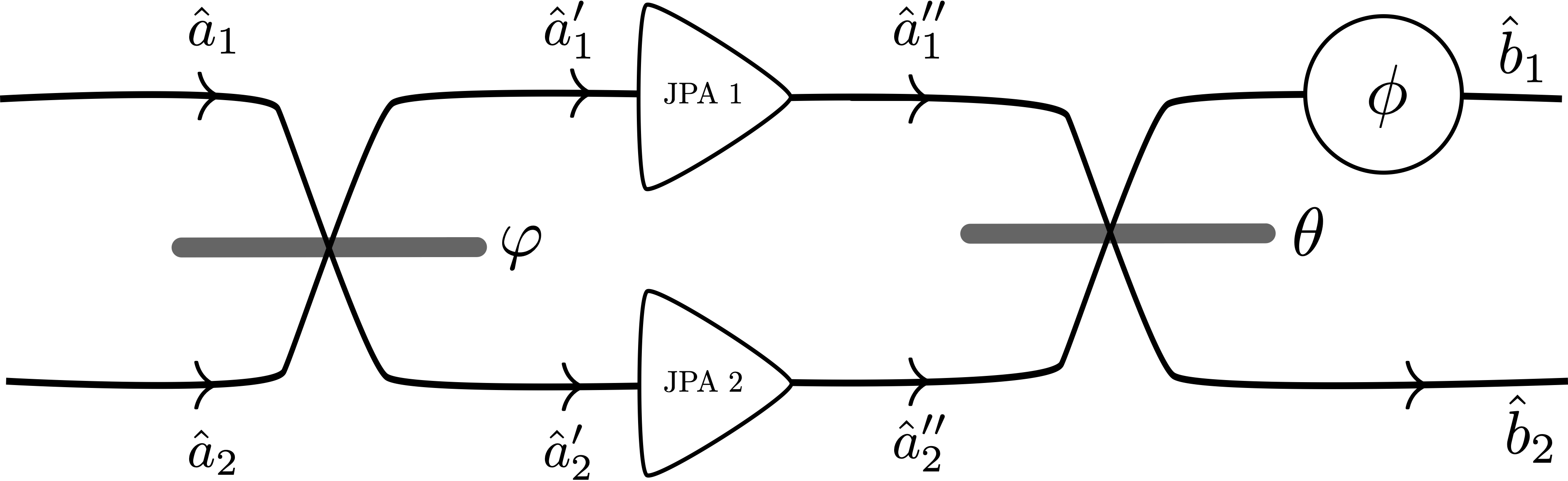}
\caption{Schematic circuit for the generation of mode $\hat{b}_1$, needed to correctly implement the optimal observable in the two-mode squeezed vacuum (TMSV) state. The original $\hat{a}_i$ modes mix at a $\varphi$ beam splitter, and the outputs go through a single-mode squeezing operator --Josephson parametric amplifier (JPA) in microwaves, with parameters $r_i$ and $\theta_i$ corresponding to squeezing and phase. Then they mix at a second beam splitter $\theta$. A phase shift $\phi$ is applied at the end, to cancel undesired terms. This scheme is  technology-independent, and could be applied to optics by replacing the JPAs with the corresponding squeezing device, \textit{e.g.} a non-linear crystal that performs spontaneous parametric down-conversion.}
\label{fig:jpas-supp}
\end{figure}

 Moreover, it is illustrative to study a possible implementation of the noiseless case, since this captures the essence of what is being measured. When $N_\text{th}\rightarrow 0 $ we have that  $\hat{O}_Q ^{\text{Lim}} =
-\mu^2\hat{a}_1^\dagger \hat{a}_1 - \hat{a}_2^\dagger \hat{a}_2 
 + \mu( \hat{a}_1^\dagger \hat{a}_2^\dagger + \hat{a}_1 \hat{a}_2)  -\nu \Id_{12}$
where $\mu^2 \equiv \left(1+1/2 N_\text{S}\right)$ and $\nu \equiv \left(1+1/4 N_\text{S}\right)$, and we have taken the limit of vanishing $N_\text{th}$. Notice that we can rewrite this observable as $\hat{b}_1^\dagger \hat{b}_1 - 1$, \textit{i.e.} implementing photon-counting on the operator $\hat{b}_1\equiv -i (\hat{a}_2^\dagger - \mu\hat{a}_1)$. This is achieved by means of the transformations captured in FIG. \ref{fig:jpas-supp}.

{ Following that scheme, we have that after the first beam splitter
\begin{align}
\begin{split}
\hat{a}_1^\prime &=  \hat{a}_1\cos\varphi + \hat{a}_2\sin\varphi \\
\hat{a}_2^\prime &=  	-\hat{a}_1\sin\varphi   + \hat{a}_2\cos\varphi ,
\end{split}\label{eq:aprimes-supp}
\end{align}
then the Josephson parametric amplifiers (JPA) --ideally squeezing operators-- produce $\hat{a}_i^{\prime \prime} = S^\dagger(r_i, \theta_i) \hat{a}_i^\prime S(r_i, \theta_i)$
where $S(r_i, \theta_i)$ is the squeezing operator, acting as
\begin{align*}
\begin{split}
\hat{a}_i^{\prime \prime}=S^\dagger(r_i, \theta_i) \hat{a}_i^\prime S(r_i, \theta_i) &= \hat{a}_i^\prime \cosh r_i  - e^{i\theta_i} \hat{a}^{\prime \dagger}_i \sinh r_i \\
{\hat{a}^{\prime \prime \dagger}}_i=S^\dagger(r_i, \theta_i) \hat{a}_i^{\prime\dagger} S(r_i, \theta_i) &=  \hat{a}_i^{\prime\dagger} \cosh r_i  - e^{-i\theta_i} \hat{a}^\prime_i \sinh r_i .
\end{split}
\end{align*}
Assuming that the phase shifter $\phi$ acts as $\hat{c} \mapsto e^{-i\phi}\hat{c}$ we find the following output modes
\begin{align*}
\begin{split}
e^{i\phi}\hat{b}_1 &= \cos \theta \left(\hat{a}_1^\prime \cosh r_1  - e^{i\theta_1} \hat{a}^{\prime \dagger}_1 \sinh r_1 \right) 
 + \sin\theta  \left(\hat{a}_2^\prime \cosh r_2  - e^{i\theta_2} \hat{a}^{\prime \dagger}_2 \sinh r_2 \right) \\
 \hat{b}_2 &= -\sin \theta \left(\hat{a}_1^\prime \cosh r_1  - e^{i\theta_1} \hat{a}^{\prime \dagger}_1 \sinh r_1 \right) 
 + \cos\theta  \left(\hat{a}_2^\prime \cosh r_2  - e^{i\theta_2} \hat{a}^{\prime \dagger}_2 \sinh r_2 \right).
\end{split}
\end{align*}
We insert \eqref{eq:aprimes-supp} in the last expression and regroup, finding
\begin{align*}
\begin{split}
e^{i\phi}\hat{b}_1 
&= \hat{a}_1 \left(   \cos\theta  \cos \varphi \cosh r_1- \sin \theta \sin \varphi  \cosh r_2 \right) \\ 
&+ \hat{a}_2 \left(   \cos\theta  \sin \varphi \cosh r_1+ \sin \theta \cos \varphi  \cosh r_2 \right) \\ 
&+ \hat{a}_1^\dagger \left(   -e^{i\theta_1}\cos\theta \cos\varphi  \sinh r_1+ e^{i\theta_2}\sin \theta  \sin \varphi \sinh r_2\right) \\ 
&+ \hat{a}_2^\dagger \left(   -e^{i\theta_1}\cos\theta \sin\varphi  \sinh r_1- e^{i\theta_2}\sin \theta  \cos \varphi \sinh r_2\right)
\end{split}
\end{align*} 
Because we want to perform photon-counting over the operator $\hat{b}_1\equiv -i (\hat{a}_2^\dagger - \mu\hat{a}_1)$, we identify:
\begin{align}
i \mu&= \cos\theta  \cos \varphi \cosh r_1- \sin \theta \sin \varphi  \cosh r_2\\
i &= e^{i\theta_1}\cos\theta \sin\varphi  \sinh r_1+ e^{i\theta_2}\sin \theta  \cos \varphi \sinh r_2.
\end{align}
}
\subsection{Optimal observable for the coherent state probe}
The optimal observable in this case is given by
$\hat{O}_C =  A \Id_{(1)}\otimes[( \hat{a}_2^\dagger - \eta_1 \sqrt{\alpha} ) (\hat{a}_2 - \eta_1 \sqrt{\alpha} ) + \frac{1}{2}]$,
where $A=1/(\eta_1-1)(1-N_\text{th}(\eta_1-1))$, and $\Id_{(1)}$ is the absence of active measurement of mode 1. 
This expression can then be put as 
$\hat{O}_C =  A\Id_{(1)}\otimes\left[\left( \hat{a}_2^\dagger - \eta_1 \sqrt{\alpha} \right) \Big(\hat{a}_2 - \eta_1 \sqrt{\alpha} \Big) + \frac{1}{2}\right]$. This operator can be experimentally performed with a displacement $D(-\eta_1 \sqrt{\alpha})$ \cite{Paris1996} and photon-counting in the resulting mode.
The interpretation is simple: because $\eta_1$ is known (it serves as a reference), there is nothing to be gained by measuring the first mode in the absence of entanglement. Moreover, the observable is separable, as one should expect, and the experimental implementation is straightforward: photon-counting in the --locally displaced-- second mode.

We have seen that both quantum and classical observables are non-Gaussian measurements, since they can be related to photon-counting, as expected in order to obtain quantum enhancement \cite{serafini}. Current photoncounters in microwave technologies can resolve up to 3 photons with an efficiency of 96\% \cite{dassonneville2020numberresolved}. Inefficiencies in the photon-counters can be accounted for with a simple model of an additional beam splitter that mixes the signal with either a vacuum or a low-tempereature thermal state. Additionally, the fact that real digital filters are not perfectly sharp  should also be accounted for in a full experimental proposal, which we leave for future work \cite{PhysRevLett.117.020502}.

\section{Conclusions}
We have proposed a novel protocol for achieving a quantum enhancement in the decision problem of whether a target's reflectivity depends or not on the frequency, using a bi-frequency, entangled probe, in the presence of  noise and losses. Crucially, our protocol needs no idler mode, avoiding the necessity of coherently storing a quantum state in a memory. 
The scaling of the quantum Fisher information (QFI) associated to the estimation problem for the entangled probe is faster than in the case of a coherent signal.
This quantum enhancement is more significant in the high reflectivity regime. Moreover, we have derived analytic expressions for the optimal observables,  which allow extraction of the maximum available information about the parameter of interest, sketching an implementation with quantum microwaves.

This information can be related to the electromagnetic response of a reflective object to changes in frequency, and, consequently, the protocol can be applied to a wide spectrum of situations. Although the results are general, we suggest two applications withing quantum microwave technology: radar physics, motivated by the atmospheric transparency window in the microwaves regime, together with the naturally noisy character of open-air  \cite{Sanz_2018, dassonneville2020numberresolved, GonzalezRaya2020, Zhang2020, munuera2020, dambach2107}; and quantum-enhanced microwave medical contrast-imaging of low penetration depth tissues, motivated not only by the non-inoizing nature of these frequencies, but also because resorting to methods that increase the precision and/or resolution without increasing the intensity of radiation is crucial in order not to heat the sample.

Our work paves the way for extensions of the protocol to accommodate both thermal effects in the input modes, and continuous-variable frequency entanglement \cite{PhysRevLett.84.5304}, where a more realistic model for a beam containing a given distribution of frequencies could be used instead of sharp, ideal bi-frequency states. It also serves as reminder that quantum enhancement provided by entanglement can survive noisy, lossy channels.

\appendix
\section{Coefficients for the optimal quantum observable}\label{app:A}
Here we give the general expressions of the coefficients of the optimal observable for the TMS state:
\begin{equation}
\hat{O}_Q = L_{11}\hat{a}_1^\dagger \hat{a}_1 + L_{22}\hat{a}_2^\dagger \hat{a}_2 + L_{12}\left( \hat{a}_1^\dagger \hat{a}_2^\dagger + \hat{a}_1 \hat{a}_2\right) + L_0 \Id_{12},
\end{equation}
{\tiny
\begin{align*}
\begin{split}
L11 &=-\frac{2 \eta _1 N_\text{S} \left(2 N_\text{S}+1\right) \left(2 N_{\text{th}}+1\right)}{-A+B-C+D}\\
L22 &= \frac{4 \eta _1 \left(2 \eta _1-1\right) N_\text{S}^2 \left(2 N_{\text{th}}+1\right)+2 N_\text{S} \left(\eta _1-2 N_{\text{th}} \left(\left(\eta _1-3\right) \eta _1+\left(\eta _1-1\right) \left(3 \eta _1-1\right) N_{\text{th}}+1\right)-1\right)+N_{\text{th}} \left(2 \left(\eta _1-1\right) N_{\text{th}} \left(\left(\eta _1-1\right) N_{\text{th}}-1\right)+1\right)}{A-B+C-D}\\
L12 &= -\frac{\sqrt{2} \sqrt{N_\text{S} \left(2 N_\text{S}+1\right)} \left(\eta _1^2 \left(N_\text{S} \left(4 N_{\text{th}}+2\right)-N_{\text{th}}^2\right)+N_{\text{th}} \left(N_{\text{th}}+1\right)\right)}{A-B+C-D}\\
L_0 &=-\frac{4 \eta _1^3 \left(N_{\text{th}}^2-2 N_\text{S} \left(2 N_{\text{th}}+1\right)\right){}^2-2 \eta _1^2 \left(6 N_{\text{th}}+5\right) F \left(N_\text{S} \left(4 N_{\text{th}}+2\right)-N_{\text{th}}^2\right)+4 \eta _1 \left(N_{\text{th}}+1\right) \left(N_\text{S}^2 \left(8 N_{\text{th}}+4\right)-2 N_\text{S} \left(N_{\text{th}} \left(6 N_{\text{th}}+5\right)+1\right)+N_{\text{th}}^2 \left(3 N_{\text{th}}+2\right)\right)+\left(2 N_{\text{th}}+3\right) G F}{E-4 \eta _1 \left(2 N_{\text{th}}+1\right) F \left(-4 N_\text{S} N_{\text{th}}+N_\text{S} \left(2 N_\text{S}-1\right)+N_{\text{th}}^2\right)-8 N_\text{S}^2 \left(3 N_{\text{th}} \left(N_{\text{th}}+1\right)+1\right)+4 N_\text{S} N_{\text{th}} \left(N_{\text{th}} \left(4 N_{\text{th}}+3\right)+1\right)-2 N_{\text{th}}^2 G},
\end{split}
\end{align*}
}%
where 
\begin{align*}
\begin{split}
A &\equiv 8 \left(\eta _1-1\right) \eta _1 N_\text{S}^3 \left(2 N_{\text{th}}+1\right)\\
B &\equiv 4 N_\text{S}^2 \left(-\eta _1+\left(\eta _1+3 \eta _1 N_{\text{th}}\right){}^2-\eta _1 N_{\text{th}} \left(10 N_{\text{th}}+7\right)+3 N_{\text{th}} \left(N_{\text{th}}+1\right)+1\right)\\
C &\equiv 2 N_\text{S} N_{\text{th}} \left(-\eta _1+N_{\text{th}} \left(\eta _1 \left(3 \eta _1-8\right)+4 \left(\eta _1-1\right) \left(2 \eta _1-1\right) N_{\text{th}}+3\right)+1\right)\\
D &\equiv N_{\text{th}}^2 \left(2 \left(\eta _1-1\right) N_{\text{th}} \left(\left(\eta _1-1\right) N_{\text{th}}-1\right)+1\right)\\
E &\equiv 4 \eta _1^2 \left(-4 N_\text{S} N_{\text{th}}+N_\text{S} \left(2 N_\text{S}-1\right)+N_{\text{th}}^2\right) \left(N_\text{S} \left(4 N_{\text{th}}+2\right)-N_{\text{th}}^2\right)\\
F &\equiv 2 N_\text{S}-N_{\text{th}}\\
G &\equiv  2 N_{\text{th}} \left(N_{\text{th}}+1\right)+1
\end{split}
\end{align*}
In the high reflectivity case $\eta_1 \rightarrow 1$ we find:
\begin{eqnarray*}
\lim_{\eta_1\rightarrow 1}L_{11} &=-\frac{2 N_\text{S} (2 N_\text{S}+1) (2 N_\text{th}+1)}{N_\text{S}^2 (8 N_\text{th} (N_\text{th}+1)+4)+4 N_\text{S} N_\text{th}^2+N_\text{th}^2}\\
\lim_{\eta_1 \rightarrow 1}L_{22} &=-\frac{4 N_\text{S} (2 N_\text{S} N_\text{th}+N_\text{S}+N_\text{th})+N_\text{th}}{N_\text{S}^2 (8 N_\text{th} (N_\text{th}+1)+4)+4 N_\text{S} N_\text{th}^2+N_\text{th}^2}\\
\lim_{\eta_1 \rightarrow 1}L_{12} &=\frac{2 \sqrt{2} \sqrt{N_\text{S} (2 N_\text{S}+1)} (N_\text{S} (4 N_\text{th}+2)+N_\text{th})}{N_\text{S}^2 (8 N_\text{th} (N_\text{th}+1)+4)+4 N_\text{S} N_\text{th}^2+N_\text{th}^2}\\
\lim_{\eta_1 \rightarrow 1}L_0 &=\frac{-2 N_\text{S} (N_\text{S} (8 N_\text{th}+4)+6 N_\text{th}+1)-3 N_\text{th}}{8 N_\text{S}^2 (2 N_\text{th} (N_\text{th}+1)+1)+8 N_\text{S} N_\text{th}^2+2 N_\text{th}^2}
\end{eqnarray*}
Additionally, as shown in the main text, in the noiseless case we get
\begin{align*}
 \hat{O}_Q ^{\text{Lim}}:=\lim_{\substack{N_\text{th}\rightarrow 0 \\ \eta_1 \rightarrow 1}} \hat{O}_Q 
 &=
-\mu^2\hat{a}_1^\dagger \hat{a}_1 - \hat{a}_2^\dagger \hat{a}_2 
+ \mu\left( \hat{a}_1^\dagger \hat{a}_2^\dagger + \hat{a}_1 \hat{a}_2\right)  -\nu \Id_{12}\\
&\equiv \hat{b}_1^\dagger \hat{b}_1 - 1,
\end{align*}
where $\mu^2 \equiv \left(1+1/2 N_\text{S}\right)$ and $\nu \equiv \left(1+1/4 N_\text{S}\right)$ and  $
\hat{b}_1\equiv -i \left(\hat{a}_2^\dagger - \mu\hat{a}_1\right)$.

 \subsection*{Acknowledgements}
The authors acknowledge the support from the EU H2020 Quantum Flagship project QMiCS (820505). MC acknowledges support from the DP-PMI and FCT (Portugal) through scholarship PD/BD/135186/2017. MC and YO thank the support from Funda\c{c}\~{a}o para a Ci\^{e}ncia e a Tecnologia (Portugal), namely through project UIDB/04540/2020, as well as from project TheBlinQC supported by the EU H2020 QuantERA ERA-NET Cofund in Quantum Technologies and by FCT (QuantERA/0001/2017). MS acknowledge financial support from Basque Government QUANTEK project from ELKARTEK program (KK-2021/00070), Spanish Ram\'on y Cajal Grant RYC-2020-030503-I and the project grant PID2021-125823NA-I00 funded by MCIN/AEI/10.13039/501100011033 and by "ERDF A way of making Europe'' and "ERDF Invest in your Future'', as well as from OpenSuperQ (820363) of the EU Flagship on Quantum Technologies, and the EU FET-Open projects Quromorphic (828826) and EPIQUS (899368).


\begin{thebibliography}{40}
\bibitem{nielsen_chuang_2010} M. A. Nielsen and I. L. Chuang. \textit{Quantum Computation and Quantum Information: 10th Anniversary Edition}. Cambridge University Press, 2010.

\bibitem{google-quantum-supremacy} F. Arute, J. Arya, R. Babbush, D. Bacon, J. C. Bardin, R. Barends, R. Biswas, S. Boixo, F. G. S. L. Brandao, H. Neven,  J. M. Martinis \textit{et al.}. Quantum supremacy using a programmable superconducting processor. Nature, 574(7779):505–510 (2019).

\bibitem{PhysRevLett.69.3598} S. L. Braunstein. Quantum limits on precision measurements of phase. Phys. Rev. Lett., {\bf 69} 3598–3601 (1992).

\bibitem{Pirandola_2018} S. Pirandola, B. R. Bardhan, T. Gehring, C. Weedbrook, and S. Lloyd. Advances in photonic quantum sensing. Nature Photonics, 12(12):724–733, (2018).

\bibitem{katori} H. Katori, Optical lattice clocks and quantum metrology, Nature Photonics  \textbf{5}, 203–210 (2011).

\bibitem{Giovannetti_2011} V. Giovannetti, S. Lloyd, and L. Maccone. Advances in quantum metrology. Nature Photonics, 5(4):222–229 (2011).

\bibitem{bongs} K. Bongs and Y. Singh, Earth-based clocks test general relativity, Nature Photonics \textbf{14}, 408–409 (2020).

\bibitem{menoret} V. Ménoret, P. Vermeulen, N. Le Moigne, S. Bonvalot, P. Bouyer, A. Landragin, and B. Desruelle, Gravity measurements below $10^{-9}$ g with a transportable absolute quantum gravimeter, Scientific Reports  \textbf{8}, 12300 (2018).


\bibitem{flury}  J. Flury, Relativistic geodesy, Journal of Physics: Conference Series 723, 012051 (2016).

\bibitem{grotti}J. Grotti et al., Geodesy and metrology with a transportable optical clock, Nature Physics \textbf{14}, 437–441 (2018).

\bibitem{zeuthen} E. Zeuthen, E. S. Polzik, and F. Ya. Khalili, Gravitational wave detection beyond the standard quantum limit using a negative-mass spin system and virtual rigidity, 
Phys. Rev. D \textbf{100}, 062004 (2019).

\bibitem{ligo} Aasi, J., Abadie, J., Abbott, B. et al., Enhanced sensitivity of the LIGO gravitational wave detector by using squeezed states of light. Nature Photonics \textbf{7}, 613–619 (2013).

\bibitem{okeke} Remote quantum clock synchronization without synchronized clocks
E. O. Ilo-Okeke, L. Tessler, J. P. Dowling, and T. Byrnes, NPJ: Quantum Information, volume 4, Article number: 40 (2018).

\bibitem{thermometry} A. De Pasquale, T. M. Stace, Quantum Thermometry (2018). In: F. Binder, L. Correa, C. Gogolin, J. Anders,  G. Adesso  (eds),  \textit{Thermodynamics in the Quantum Regime}. Fundamental Theories of Physics, vol 195. Springer, Cham. 

\bibitem{bowen2013} Taylor, M., Janousek, J., Daria, V. et al., Biological measurement beyond the quantum limit, Nature Photonics \textbf{7}, 229–233 (2013).

\bibitem{omar} M. H. Mohammady, H. Choi, M. E. Trusheim, A. Bayat, D. Englund, Y. Omar, Low-control and robust quantum refrigerator and applications with electronic spins in diamond, Phys. Rev. A \textbf{97}, 042124 (2018).


\bibitem{plenio} J. Cai, F. Jelezko, and M. B. Plenio, Hybrid sensors based on colour centres in diamond and piezoactive layers, Nature Communications  \textbf{5}: 4065 (2014).

\bibitem{biosensors} F. Dolde et al., Electric-field sensing using single diamond spins,
Nature Physics  \textbf{7}, 459–463 (2011).

\bibitem{jensen} K. Jensen et al.,  Non-invasive detection of animal nerve impulses with an atomic magnetometer operating near quantum limited sensitivity,  Scientific Reports  \textbf{6}, 29638 (2016).

\bibitem{milburn} S. Bose et al., A Spin Entanglement Witness for Quantum Gravity, Phys. Rev. Lett. \textbf{119}, 240401 (2017).

\bibitem{lee} S. Lee, Y. S. Ihn, and Z. Kim, Optimal entangled coherent states in lossy quantum-enhanced metrology, Phys. Rev. A \textbf{101}, 012332 (2020).

\bibitem{changhun} C. Oh, S. Lee, H. Nha, and H. Jeong, Practical resources and measurements for lossy optical quantum metrology, Phys. Rev. A \textbf{96}, 062304 (2017).

\bibitem{zhang} X. Zhang, Y. Yang, and X. Wang, Lossy quantum-optical metrology with squeezed states, Phys. Rev. A \textbf{88}, 013838 (2013).

\bibitem{zhang2}Y. M. Zhang, X. W. Li, W. Yang, and G. R. Jin,  Quantum Fisher information of entangled coherent states in the presence of photon loss, Phys. Rev. A \textbf{88}, 043832 (2013).

\bibitem{guta2012} R. Demkowicz-Dobrzański, J. Kołodyński, and  M. Guţă, The elusive Heisenberg limit in quantum-enhanced metrology. Nat Commun \textbf{3}, 1063 (2012).

\bibitem{davidovich2012}
B. M. Escher, L. Davidovich, N. Zagury, and R. L. de Matos Filho, Quantum Metrological Limits via a Variational Approach, 
Phys. Rev. Lett. \textbf{109}, 190404 (2012).

\bibitem{acin2013}
R. Chaves, J. B. Brask, M. Markiewicz, J. Kołodyński, and A. Acín, Noisy Metrology beyond the Standard Quantum Limit,
Phys. Rev. Lett. \textbf{111}, 120401 (2013).

\bibitem{sekatski} P. Sekatski et al., Dynamical decoupling leads to improved scaling in noisy quantum metrology,  New J. Phys. \textbf{18} 073034 (2016).

\bibitem{sekatski2} P. Sekatski et al., Quantum metrology with full and fast quantum control,  Quantum \textbf{1}, 27 (2017).

\bibitem{huelga} J. F. Haase, A. Smirne, J. Ko\l ody\' nski, R. Demkowicz-Dobrza\' nski, and S. F. Huelga, Precision Limits in Quantum Metrology with Open Quantum Systems, Quantum Measurements and Quantum Metrology \textbf{5}, 13 (2018)

\bibitem{friis} N. Friis, Flexible resources for quantum metrology, \textit{et al.}  New J. Phys. \textbf{19}, 063044 (2017).

\bibitem{huelga2014} J. Jeske, J. H. Cole, and S. F.  Huelga, Quantum metrology subject to spatially correlated Markovian noise: restoring the Heisenberg limit, 2014 New J. Phys. \textbf{16} 073039 (2014).

\bibitem{sekatski2017} R. Demkowicz-Dobrzański, J. Czajkowski, and P. Sekatski, Adaptive Quantum Metrology under General Markovian Noise, Phys. Rev. X \textbf{7}, 041009 (2017).

\bibitem{muschik2017}F. Reiter, A. S. Sørensen, P. Zoller et al., Dissipative quantum error correction and application to quantum sensing with trapped ions. Nature Communications \textbf{8}, 1822 (2017).

\bibitem{Lloyd1463} S. Lloyd, Enhanced Sensitivity of Photodetection via Quantum Illumination, Science {\bf 321}, 5895 (2008).

\bibitem{PhysRevLett.101.253601}  S. Tan, B. I. Erkmen, V. Giovannetti, S. Guha, S. Lloyd, L. Maccone, S. Pirandola, and J. H. Shapiro, Quantum Illumination with Gaussian States, Phys. Rev. Lett.  {\bf 101}, 253601 (2008).

\bibitem{guha2009} S. Guha and B. I. Erkmen, Gaussian-state quantum-illumination receivers for target detection, Phys. Rev. A \textbf{80}, 052310 (2009)

\bibitem{shapiroStory} J. H. Shapiro, The Quantum Illumination Story, IEEE Aerospace and Electronic Systems Magazine, \textbf{35}, no. 4, pp. 8-20 (2020).

\bibitem{alsing2019}  S. Ray, J. Schneeloch, C. C. Tison, P. M. Alsing, The maximum advantage of quantum illumination, Phys. Rev. A \textbf{100}, 012327 (2019).

\bibitem{pirandolaMicrowave} S. Barzanjeh, S. Guha, C. Weedbrook, D. Vitali, J. H. Shapiro, and S. Pirandola, Microwave Quantum Illumination, Phys. Rev. Lett. \textbf{114}, 080503 (2015).

\bibitem{shabir} S. Barzanjeh, S. Pirandola, D. Vitali, J. M. Fink, Microwave quantum illumination using a digital receiver, Science Advances  Vol. 6, no. 19 (2020).

\bibitem{borre} G. De Palma, J. Borregaard, The minimum error probability of quantum illumination, Phys. Rev. A \textbf{98}, 012101 (2018).

\bibitem{genovese} E. D. Lopaeva, I. Ruo Berchera, I. P. Degiovanni, S. Olivares, G. Brida, M. Genovese, Experimental realisation of quantum illumination, 	Phys. Rev. Lett. \textbf{110}, 153603 (2013)

 \bibitem{gregory}T. Gregory, P.-A. Moreau, E. Toninelli, and M. J. Padgett, 
Imaging through noise with quantum illumination, Science Advances, Vol. 6, no. 6, eaay2652 (2020).

\bibitem{cai2021} Q. Cai, J. Liao, B. Shen, G. Guo, and Q. Zhou, Microwave quantum illumination via cavity magnonics, Phys. Rev. A 103, 052419 (2021)

\bibitem{lanzagorta} M. Lanzagorta and J. Uhlmann, \textit{Quantum Radar} (Morgan
and Claypool, San Rafael, CA, 2012), p. 66


\bibitem{macconeRadar} L. Maccone and C. Ren,  Quantum Radar, Phys. Rev. Lett. \textbf{124}, 200503 (2020).


\bibitem{Zhuang2021} Q. Zhuang, Quantum Ranging with Gaussian Entanglement, Phys. Rev. Lett. \textbf{126}, 240501 (2021).

\bibitem{Sanz_2017} M. Sanz, U. Las Heras, J. J. García-Ripoll, E. Solano, and R. Di Candia. Quantum estimation methods for quantum illumination. Phys. Rev. Lett., {\bf 118}, 070803 (2017).

\bibitem{williamson} J. Williamson,  On the algebraic problem concerning the normal forms of linear dynamical systems, Am. J. Math. \textbf{58},  141 (1936).

\bibitem{personick} S. Personick, Application of quantum estimation theory to analog communication over quantum channels, IEEE Trans. Inf. Theory \textbf{17}, 240-6 (1971).

\bibitem{marcin} M. Jarzyna and R. Demkowicz-Dobrzański, True precision limits in quantum metrology,  New J. Phys. \textbf{17}, 013010 (2015).

\bibitem{morelli} S. Morelli et al., Bayesian parameter estimation using Gaussian states and measurements, Quantum Sci. Technol. \textbf{6}, 025018 (2021).

\bibitem{pi} 
W. Górecki, R. Demkowicz-Dobrzański, H. M. Wiseman, and D. W. Berry, $\pi$-Corrected Heisenberg Limit, Phys. Rev. Lett. 124, 030501 (2020).

\bibitem{rubio1} J. Rubio, P. Knott, and J. Dunningham, Non-asymptotic analysis of quantum metrology protocols beyond the Cramér-Rao bound,  J. Phys. Commun.
\textbf{2} 015027 (2018). 

\bibitem{rubio2}  J. Rubio and J.
Dunningham, Quantum metrology in the presence of limited data,  New J. Phys. \textbf{21} 043037 (2019).

\bibitem{rubio3} J. Rubio
and J. Dunningham, Bayesian multi-parameter quantum metrology with limited data, Phys. Rev. A \textbf{101}, 032114 (2020).

\bibitem{kay1993}
S. Kay, Fundamentals of Statistical Signal Processing: Estimation Theory (Prentice Hall, 1993).

\bibitem{demko2015}
R. Demkowicz-Dobrza\' nski, M. Jarzyna, and J. Kołody\' nski,
Quantum Limits in Optical Interferometry, Progress in Optics
60, 345 (2015).

\bibitem{arvind} D. B. Arvind, N. Mukunda, and R. Simon,  The real symplectic groups in quantum mechanics and optics, Pramana \textbf{45} 471–97 (1995).

\bibitem{simon} R. Simon, V. Srinivasan, and S.K. Chaturvedi, Congruences and Canonical
Forms for a Positive Matrix: Application to the Schweinler-Wigner Extremum Principle, J. Math. Phys., \textbf{40}:7 (1998).

\bibitem{holevo} 
A. Holevo, {\it Probabilistic and Statistical Aspects of Quantum Theory}, ISBN: 978-88-7642-375-8 (Springer, 2011).

\bibitem{helstrom} 
C. W. Helstrom, Quantum Detection and Estimation Theory, Journal of Statistical Physics {\bf 1}, 231 (1969).

\bibitem{caves}  S. L. Braunstein and C. M. Caves, Statistical Distance and the Geometry of Quantum States, Phys. Rev. Lett. \textbf{72}, 3439 (1994).

\bibitem{doi:10.1142/S0219749909004839} M. G. A. Paris. Quantum estimation for quantum technology. International Journal of Quantum Information, 07(supp01):125–137 (2009).

\bibitem{cramer}
H. Cramér, Mathematical methods of statistics, Vol.43 (Princeton university press, 1999).

\bibitem{rao} C. R. Rao, Information and the accuracy attainable in the estimation of statistical parameters, in Breakthroughs in statistics (Springer, 1992) pp. 235–247.

\bibitem{parisBook}M. G. A. Paris, J. Rehacek (Eds.), \textit{Quantum State Estimation}, Springer-Verlag Berlin Heidelberg, 2004.

\bibitem{footnote1} In a multiparameter scenario the quantum Fisher information is actually a matrix.


\bibitem{parisOlivares} A. Ferraro, S. Olivares, M. G. A. Paris, \textit{Gaussian States in Quantum Information}, Bibliopolis (2005).


\bibitem{adesso} G. Adesso, S. Ragy, A. R. Lee, Continuous variable quantum information: Gaussian states and beyond, Open Syst. Inf. Dyn. \textbf{21}, 1440001 (2014).

\bibitem{olivares} S. Olivares, Quantum optics in the phase space - A tutorial on Gaussian states, Eur. Phys. J. Special Topics \textbf{203}, 3-24 (2012).

\bibitem{serafini}
{A. Serafini, \textit{Quantum Continuous Variables: A Primer of Theoretical Methods} (CRC Press, 2017).}

\bibitem{Safranek_2018} D. Šafránek. Estimation of Gaussian quantum states. Journal of Physics A: Mathematical and Theoretical, 52(3):035304 (2018).

\bibitem{_afr_nek_2015} D. Šafránek, A. R. Lee, and I. Fuentes. Quantum parameter estimation using multi-mode Gaussian states. New Journal of Physics, 17(7):073016 (2015).

\bibitem{pinel}  O. Pinel, J. Fade, D. Braun, P. Jian, N. Treps, and
C. Fabre, Ultimate sensitivity of precision measurements with intense Gaussian quantum light: A multimodal approach, Phys. Rev. A \textbf{85}, 010101 (2012).

\bibitem{friis2015} N. Friis, M. Skotiniotis, I. Fuentes, and W. Dür, Heisenberg scaling in Gaussian quantum metrology, Phys. Rev. A \textbf{92}, 022106 (2015).

\bibitem{pinel2013} 
O. Pinel, P. Jian, N. Treps, C. Fabre, and D. Braun, Quantum parameter estimation using general single-mode Gaussian states, 
Phys. Rev. A 88, 040102(R) (2013).

\bibitem{monras} A. Monras, Phase space formalism for quantum estimation of Gaussian states, 	arXiv:1303.3682 (2013).

\bibitem{jiang2013} Z. Jiang, Quantum Fisher information for states in exponential form, Phys. Rev. A 89, 032128 (2013).

\bibitem{marian2016} P. Marian and T. A. Marian,
Quantum Fisher information on two manifolds of two-mode Gaussian states,  Phys. Rev. A \textbf{93}, 052330 (2016).

\bibitem{gao}Y. Gao and H. Lee, Bounds on quantum multiple-parameter estimation with Gaussian state, Eur. Phys. J. D \textbf{68}, 347 (2014).

\bibitem{nichols} R. Nichols, P. Liuzzo-Scorpo, P. A. Knott, and G. Adesso, Multiparameter Gaussian quantum metrology, Phys. Rev. A \textbf{98}, 012114 (2018).

\bibitem{banchi}  L. Banchi, S. L. Braunstein, and S. Pirandola, Quantum Fidelity for Arbitrary Gaussian States, Phys. Rev. Lett. \textbf{115}, 260501 (2015).





\bibitem{Sanz_2018} M. Sanz, K. G. Fedorov, F. Deppe, and E. Solano. Challenges in open-air microwave quantum communication and sensing. 2018 IEEE Conference on Antenna Measurements \& Applications (CAMA) (2018).

\bibitem{PhysRevLett.107.113601} C. Eichler, D. Bozyigit, C. Lang, M. Baur, L. Steffen, J. M. Fink, S. Filipp, and A. Wallraff. Observation of two-mode squeezing in the microwave frequency domain. Phys. Rev. Lett., {\bf 107} 113601 (2011).

\bibitem {PhysRevLett.109.183901} E. Flurin, N. Roch, F. Mallet, M. H. Devoret, and B. Huard. Generating entangled microwave radiation over two transmission lines. Phys. Rev. Lett., {\bf 109} 183901 (2012).

\bibitem {RevModPhys.77.513} S. L. Braunstein and P. van Loock. Quantum information with continuous variables. Rev. Mod. Phys., 77:513–577 (2005).


\bibitem {casariego2022} M. Casariego, E. Z. Cruzeiro, S. Gheradini \textit{et al.}, 	arXiv:2205.11424 [quant-ph] (2022).

\bibitem {gonzalezraya2022} T. Gonzalez-Raya, M. Casariego, F. Fesquet, \textit{et al.}, 		arXiv:2203.07295 [quant-ph] (2022).

\bibitem{footnote2} Note that $N_\text{S}$ is defined as the expected photon number in a \textit{single} mode, meaning that the actual photon number in the signal is $2N_\text{S}$.

\bibitem{footnote3} In a state language this translates to the total unitary being the tensor product of two beam splitters: ${U}_T(\eta_1,\eta_2)= {U}(\eta_1) \otimes {U}(\eta_2)$, with ${U}(\eta_i)\equiv \exp \left[ \arcsin\sqrt{\eta(\omega_i)}(\hat{s}^\dagger_i \hat{b}_i-\hat{s}_i \hat{b}^\dagger_i)\right]$.

\bibitem{cloaking}  U. Las Heras, R. Di Candia, K. G. Fedorov, F. Deppe, M. Sanz, E. Solano. Quantum illumination reveals phase-shift inducing cloaking. Scientific Reports \textbf{7}, 9333 (2017).

\bibitem{TMST} J. Oz-Vogt , A. Mann, and  M. Revzen.  Thermal Coherent States and Thermal Squeezed States, Journal of Modern Optics, 38:12, 2339-2347, (1991).

\bibitem{finite-time}  K. G. Fedorov, S. Pogorzalek, U. Las Heras, M. Sanz, P. Yard, P. Eder, M. Fischer, J. Goetz, E. Xie, K. Inomata, Y. Nakamura, R. Di Candia, E. Solano, A. Marx, F. Deppe, R. Gross. Finite-time quantum entanglement in propagating squeezed microwaves. Sci Rep \textbf{8}, 6416 (2018).

\bibitem{Paris1996} M. G. A. Paris, Displacement operator by beam splitter, Phys. Lett. A 217, 78 (1996).

\bibitem{dassonneville2020numberresolved} R. Dassonneville, R. Assouly, T. Peronnin, P. Rouchon, and B. Huard. Number-resolved photocounter for propagating microwave mode, Phys. Rev. Applied \textbf{14}, 044022 (2020).

\bibitem{PhysRevLett.117.020502} K. G. Fedorov, L. Zhong, S. Pogorzalek, P. Eder, M. Fischer, J. Goetz, E. Xie, F.Wulschner, K. Inomata, T. Yamamoto, Y. Nakamura, R. Di Candia, U. Las Heras, M. Sanz, E. Solano, E. P. Menzel, F. Deppe, A. Marx, and R. Gross. Displacement of propagating squeezed microwave states. Phys. Rev. Lett., {\bf 117} 020502 (2016).


\bibitem{GonzalezRaya2020}
T. González-Raya and M. Sanz
Coplanar Antenna Design for Microwave Entangled Signals Propagating in Open Air, Quantum {\bf 6}, 783 (2022).

\bibitem{Zhang2020} J. Zhang, T .Li, R. Kokkoniemi, C. Yan, W. Liu, M. Partanen, K. Y. Tan, M. He, L. Ji, L. Grönberg, M. Möttönen,
Broadband Tunable Phase Shifter For Microwaves, arXiv:2003.01356.


\bibitem{munuera2020} C. Munuera-Javaloy, I. Arrazola, E. Solano, and J. Casanova,
Double quantum magnetometry at large static magnetic fields,
Phys. Rev. B \textbf{101}, 104411 (2020).

\bibitem{dambach2107}
S. Dambach,  B. Kubala, and J. Ankerhold, Generating entangled quantum microwaves in a Josephson-photonics device, New J. Phys. \textbf{19}, 023027 (2017).

\bibitem{PhysRevLett.84.5304} C. K. Law, I. A. Walmsley, and J. H. Eberly. Continuous frequency entanglement: Effective finite Hilbert space and entropy control. Phys. Rev. Lett., {\bf 84} 5304–5307 (2000).

\end{thebibliography}
\end{document}